\documentclass[preprint,showpacs,amsmath,amssymb,pre,superscriptaddress,floatfix]{revtex4-1}

\usepackage{graphicx}
\usepackage{dcolumn}
\usepackage{bm}
\usepackage{amssymb}
\usepackage{multirow}

\usepackage{braket}

\begin{document}

\title{Analysis of resonant population transfer in time-dependent elliptical quantum billiards}

\date{\today}


\author{Jakob Liss}
\affiliation{Zentrum f\"ur Optische Quantentechnologien, Universit\"at Hamburg, Luruper Chaussee 149, 22761 Hamburg, Germany}%
\author{Benno Liebchen}
\email[]{Benno.Liebchen@physnet.uni-hamburg.de}
\affiliation{Zentrum f\"ur Optische Quantentechnologien, Universit\"at Hamburg, Luruper Chaussee 149, 22761 Hamburg, Germany}%
\author{Peter Schmelcher}
\email[]{Peter.Schmelcher@physnet.uni-hamburg.de}
\affiliation{Zentrum f\"ur Optische Quantentechnologien, Universit\"at Hamburg, Luruper Chaussee 149, 22761 Hamburg, Germany}%

\begin{abstract}
A Fermi's Golden Rule for population transfer between instantaneous eigenstates of elliptical quantum billiards with oscillating boundaries
is derived. Thereby, both the occurrence of the recently observed resonant population transfer between instantaneous eigenstates [F. Lenz et al. New J. Phys.,  {\bf 13}, 103019, 2011]
and the empirical criterion stating that these transitions occur when the driving frequency matches
the mean difference of the latter are explained. 
As a second main result a criterion judging which resonances are resolvable in a
corresponding experiment of certain duration is provided. 
Our analysis is complemented by numerical simulations for three different driving laws. The corresponding resonance spectra 
are in agreement with the predictions of both criteria.
\end{abstract}

\maketitle

\section{Introduction}
Classical driven billiards of varying geometry have been subject to intensive research during the last years 
\cite{Loskutov1,Loskutov2,Leonel1,Leonel2,Leonel3,Leonel4,Leonel5,Matrasulov1,ClEllBill1,ClEllBill2,ExpFermiOrig,BennoExpFermi,LRA}.
While billiards are, in general, important models to study aspects of nonlinear dynamics, semi-classics or (quantum) chaos \cite{stockmann,reichl}, driven billiards additionally
facilitate the study of non-equilibrium dynamics.
As one of the key topics concerning driven billiards, Fermi acceleration (FA) and the related conditions for its occurrence have gained much attention 
\cite{Loskutov1,ExpFermiOrig,BennoExpFermi,Leonel3,Leonel4,Leonel5,ClEllBill1,ClEllBill2,LRA}. 
FA describes the unbounded growth of energy of particles that repeatedly interact with a time-dependent potential that is usually modeled by a moving billiard boundary 
and was originally proposed by Enrico Fermi as a possible mechanism to explain high-energetic cosmic radiation \cite{Fermi}.
The infamous Fermi-Ulam model (FUM) is basically a one-dimensional billiard with a moving boundary and it was found that FA is present in the FUM only 
for non-smooth driving laws \cite{liebermann}.
The general conditions for the occurrence of FA are still under debate.
Originally, it was assumed that a sufficient condition for the occurrence of FA in a driven two-dimensional 
billiard is the presence of chaotic regions in the phase space of the 
corresponding static billiard \cite{LRA}.
However, it turned out, that driving an ovally shaped billiard which has a 
mixed phase space in a certain mode does not lead to FA \cite{PhdLenz}.
On the other hand, it was shown that the classical driven elliptical billiard does show FA although its static counterpart is completely integrable 
\cite{ClEllBill1, ClEllBill2}.
Furthermore, while correlated motion suppresses FA for smooth driving in the FUM, it was found that correlations can even cause
exponential FA for smooth driving laws in a related two-dimensional model \cite{ExpFermiOrig,BennoExpFermi}.

Although it is known that periodically driven quantum billiards with a discrete Floquet spectrum can not exhibit FA \cite{BoundedEnergy}, 
it is natural to complement the study of the classical dynamics of a system by analyzing its quantum behavior.
While one finds many studies to the quantum version of the one-dimensional FUM (see \cite{QFUM1,QFUM2,QFUM3} and references therein), literature is very sparse on driven 
quantum billiards of higher dimensions \cite{Flrespop,DQB1,DQB2}.  

In particular, \cite{Flrespop} presents a method to solve the time-dependent elliptical quantum billiard.
The main result was the numerical observation of resonances in the population transfer probability between instantaneous energy eigenstates.
These transitions could be reproduced in an effective Rabi-model and captured by a criterion stating  
that resonances occur whenever the difference of corresponding time-averaged energy eigenvalues matches an integer multiple of the driving frequency.
However, an explanation for this criterion was not given in \cite{Flrespop}.

Here, we develop a systematic perturbative analysis of population transfer for the system analyzed in \cite{Flrespop} and a generalized
driving law. In this framework a Fermi's Golden Rule \cite{goldenRule} is derived for elliptical quantum billiards with oscillating boundaries 
which explains the key observations in \cite{Flrespop}, i.e. the occurrence of resonant population transfer between instantaneous eigenstates
and the empirical criterion relating these resonances with the spectrum of instantaneous eigenstates and the driving frequency. 
As a second major result, we provide a criterion to decide whether a predicted resonance can be resolved in a possible 'experiment' of a certain duration.
Finally, the numerical studies in \cite{Flrespop} are complemented by a corresponding analysis of further driving laws. 
The predictions derived within our perturbative analysis will be shown to provide a perfect agreement with the numerical results in all cases. 

This work is structured as follows: 
Chapter \ref{ch:setup} provides a short summary of the solution of the time-dependent elliptical quantum billiard
as developed in \cite{Flrespop},
followed by transformations that bring the Schr\"odinger equation into a form 
being convenient for the application of time-dependent perturbation theory.
In chapter \ref{ch:perturbative_analysis}, we will calculate the transition rate between two instantaneous eigenstates per unit time is calculated in first order perturbation theory
and an approximate population dynamics in the near-resonant case is derived. We find a criterion for the resolvability of predicted resonances in a possible experiment of certain duration.
Finally, in chapter \ref{ch:numerical_results} we present and analyze numerical results for three different periodic driving laws.

\section{Time-dependent elliptical billiard and its analytic treatment}

In the following, we first summarize our approach to a numerical solution of the time-dependent Schrödinger equation of the elliptical billiard as presented in \cite{Flrespop}.
We will transform the time-dependent Schr\"odinger equation (TDSE) into a convenient form and finally develop a perturbative approach for the periodically driven billiard in the second part of this chapter.

\subsection{Setup}
\label{ch:setup}

A wave function $\Psi(\vec{x},t)$ in a driven elliptical billiard obeying the TDSE
\begin{equation}
 \dot \imath \hbar \partial_t \Psi(\vec{x},t) = -\frac{\hbar^2}{2 \mu} \Delta \Psi(\vec{x},t) \textrm{,}
\label{Eq:SE1}
\end{equation}
is subject to Dirichlet boundary conditions, $\Psi(\vec{x},t)|_{\partial B} = 0$, on a boundary $\partial B$ of elliptical shape:
\begin{equation}
 \partial B = \left\{\vec{x} = (x,y)^{\intercal} \in \mathbb{R}^2 \left| \frac{x^2}{a^2} + \frac{y^2}{b^2} = 1\right.\right\} \textrm{.}
\end{equation}
Here, the semi-axes of the elliptical boundary, $a$ and $b$, are assumed to be arbitrary smooth functions of time, i.e. $a=a(t)$ and $b=b(t)$. 

The time-dependent boundary conditions can be handled by a coordinate transformation \cite{Flrespop},
\begin{equation}
 \rho_t: \left(\begin{array}{c}
  x\\
  y
 \end{array}\right) \mapsto
 \left(\begin{array}{c}
  \eta\\
  \xi
 \end{array}\right) =
 \left(\begin{array}{cc}
  \frac{1}{a(t)} & 0\\
  0 & \frac{1}{b(t)}
 \end{array}\right)
 \left(\begin{array}{c}
  x\\
  y
 \end{array}\right)
\label{Eq:coordtrafo}
\end{equation}
that maps the time-dependent elliptical boundary onto a static boundary of the shape of a unit circle.
Applying (\ref{Eq:coordtrafo}) to (\ref{Eq:SE1}) together with a unitary transformation,
\begin{equation}
 U(x,y,t) = \exp\left(-\frac{\dot \imath \mu}{2 \hbar}\left(\frac{\dot a(t) x^2}{a(t)} + \frac{\dot b(t) y^2}{b(t)}\right)\right) \textrm{,}
\label{Eq:U}
\end{equation}
and extracting a volume-dependent prefactor $\sqrt{a(t) b(t)}$ from the wave function $\Psi(\vec{x},t)$, we are led to an effective SE
\begin{equation}
 \dot \imath \hbar \partial_t \Lambda(\eta,\xi,t) = H^{e}(\eta,\xi,t) \Lambda(\eta,\xi,t) \textrm{,}
\label{Eq:SE2}
\end{equation}
where the effective Hamiltonian $H^{e}$ contains time derivatives of the prefactor $\sqrt{a(t) b(t)}$ and of the unitary transformation $U$ of the left-hand side of the TDSE:
\begin{equation}
 H^{e}(\eta,\xi,t) = \frac{-\hbar^2}{2 \mu} \left(\frac{1}{a^2(t)}\frac{\partial^2}{\partial \eta^2} + \frac{1}{b^2(t)}\frac{\partial^2}{\partial \xi^2} \right) 
+ \frac{1}{2} \mu \left(a(t)\ddot a(t) \eta^2 + b(t)\ddot b(t) \xi^2\right) \textrm{.}
\label{Eq:effective_Hamiltonian}
\end{equation}
The introduction of the unitary transformation (\ref{Eq:U}) ensures that $H^{e}$ is Hermitian.
Due to the extracted prefactor $\sqrt{a(t) b(t)}$, the effective wave function 
\begin{equation}
 \Lambda(\eta,\xi,t) := \sqrt{a(t)b(t)} \ U(\rho_t^{-1}(\eta,\xi),t) \Psi(\rho_t^{-1}(\eta,\xi),t)
\end{equation}
is normalized to $1$ on the domain boundary of the unit circle $C := \left\{\vec{x} = (x,y)^{\intercal} \in \mathbb{R}^2 \left| x^2 + y^2 \leq 1\right.\right\}$ and 
the coordinate transformation (\ref{Eq:coordtrafo}) makes $\Lambda$ subject to the Dirichlet boundary condition $\Psi(\vec{x},t)|_{\partial C} = 0$.
The reader is referred to \cite{Flrespop} for a similar, more detailed derivation of the effective Hamiltonian and equations of motion.

A complete set of orthonormal functions on $C$ is given by the eigenfunctions of the static circular billiard \cite{circular_billiard1,circular_billiard2}:
\begin{equation}
 \Phi_{n,m}(\rho,\phi) = \frac{1}{\sqrt{\pi} J_{m+1}(k_{m,n})} J_{m}(k_{m,n} \rho) e^{\dot \imath m \phi} \textrm{.}
\label{Eq:eigenfunctions_circular_billiard}
\end{equation}
$\rho$ and $\phi$ can be calculated from $\eta$ and $\xi$ by $\eta = \rho \cos \phi$ and $\xi = \rho \sin \phi$.
$J_m$ is the cylindrical Bessel function of order $m$ and $k_{m,n}$ is its $n$-th root. $n$ and $m$ are called radial, resp. angular, quantum number for obvious reasons.
If we expand the effective wave function $\Lambda$ in terms of the eigenfunctions of the static circular billiard, the effective SE (\ref{Eq:SE2}) becomes a linear homogeneous ordinary differential 
equation of first order in time and can, thus, be solved numerically by standard methods \cite{Flrespop}.

A main result of \cite{Flrespop} was the observation of resonant population transfer between so-called instantaneous eigenstates of
\begin{equation}
  H_M = \frac{-\hbar^2}{2 \mu} \left(\frac{1}{a^2(t)}\frac{\partial^2}{\partial \eta^2} + \frac{1}{b^2(t)}\frac{\partial^2}{\partial \xi^2} \right) \textrm{.}
\label{Eq:Mathieu_Hamiltonian}
\end{equation}
We understand instantaneous eigenstates as follows: 
The semi-axes $a$ and $b$ are parameters of $H_M$ that change in time.
If we evolve our system solely by $H_M$ in the SE, start the system in an initial state that corresponds to an eigenstate of $H_M$ at $t=0$ and change $a$ and $b$ sufficiently slowly, then
we define the instantaneous eigenstate of $H_M$ at time $t$ as the time-evolved wave function of the system at time $t$ in accordance to the adiabatic theorem of quantum mechanics \cite{adiabatensatz}. 

The Hamiltonian $H_M$ (\ref{Eq:Mathieu_Hamiltonian}) is part of the effective Hamiltonian $H^e$ (\ref{Eq:effective_Hamiltonian}).
Its complementary part is
\begin{equation}
 H_F = H^e - H_M = \frac{1}{2} \mu \left(a(t)\ddot a(t) \eta^2 + b(t)\ddot b(t) \xi^2\right) \textrm{.}
\label{Eq:coupling_hamiltonian}
\end{equation}
Population transfer between instantaneous eigenstates of $H_M$ takes place by two different mechanisms in the billiard.
First, as $a$ and $b$ are of course not changed sufficiently slowly, diabatic population transfer between the instantaneous eigenstates of $H_M$ will take place.
Additionally, the Hamiltonian $H_F$ triggers population transfer as it is non-diagonal in the basis set of instantaneous eigenstates of $H_M$. 

Introducing the volume of the elliptical billiard, $V(t) = a(t)b(t) \textrm{,}$ and the ratio of the semi-axes, $r(t) = b(t)/a(t)$, $H_M$ can be rewritten in the much more convenient form
\begin{equation}
 H_M = \frac{\hbar^2}{\mu V(t)} M(r(t)) \textrm{,}
\end{equation}
where we call
\begin{equation}
 M(r) := -\frac{1}{2}\left(r \frac{\partial^2}{\partial \eta^2} + \frac{1}{r}\frac{\partial^2}{\partial \xi^2}\right)
\label{Eq:Mathieu_op}
\end{equation}
the Mathieu operator as its eigenfunctions are just ordinary and modified Mathieu functions as they appear in the solutions of the static elliptical billiard.
If we label the eigenstates of $M(r)$ by $\Ket{n;r}$ with eigenvalue $q_n(r)$, $\Ket{n;r(t)}$ are, of course, the instantaneous eigenstates of $H_M$ and
$E_n(t) = \frac{\hbar^2}{\mu V(t)} q_n(r(t))$ the corresponding instantaneous eigenvalues of $H_M$, i.e. we have
\begin{equation}
 H_M = \sum_{n = 1}^{\infty} \Ket{n;r(t)}\frac{\hbar^2 q_n(r(t))}{\mu V(t)}\Bra{n;r(t)} \textrm{.}
\label{Eq:Mathieu_Ham_diagonal_form}
\end{equation}
Note that $M(r)$ is invariant upon sign-change of $\eta$ and $\xi$.
One can therefore choose its eigenstates $\Ket{n;r}$ such that they are also eigenstates of the parity operators that change the sign of $\eta$ or $\xi$.
In this context, we will refer to $\Ket{n;r}$ having even or odd $\eta$-, resp. $\xi$-, parity.
Note that the effective Hamiltonian $H^e$ (\ref{Eq:effective_Hamiltonian}) is also invariant upon sign-change of $\eta$ and $\xi$ and, consequently, only couples instantaneous eigenstates that have the same
$\eta$- and $\xi$- parity.
The Hilbert space therefore splits into four uncoupled Hilbert subspaces.

We choose the following ansatz for the effective wave function $\Lambda$,
\begin{equation}
 \Ket{\Lambda(t)} = \sum_n c_n(t) e^{-\frac{\dot \imath}{\hbar} \phi_n(t)} \Ket{n;r(t)} \textrm{,}
\end{equation}
with time-dependent expansion coefficients $c_n(t)$ and
\begin{equation}
 \phi_n(t) := \int_0^t dt' E_n(t') = \int_0^t dt' \frac{\hbar^2}{\mu V(t')} q_n(r(t'))
\label{Eq:phase_dim}
\end{equation}
being the time-integrated instantaneous eigenvalues of $H_M$.
If we put this ansatz into the the SE (\ref{Eq:SE2}) and note that $ \partial_t \Ket{n;r(t)} = \dot r (\partial_r \Ket{n;r})|_{r = r(t)} \equiv \dot r \partial_r \Ket{n;r(t)} \textrm{,}$
we get a SE for the coefficients $c_n(t)$:
\begin{eqnarray}
 \dot \imath \hbar \partial_t c_n(t) &=& \sum_m c_m(t) e^{-\frac{\dot \imath}{\hbar} (\phi_m(t)-\phi_n(t))} \left(\Braket{n;r(t)|H_F(t)|m;r(t)} - \dot \imath \hbar \dot r(t) \Braket{n;r(t)|\partial_r|m;r(t)} \right) \nonumber\\
\label{Eq:SE_coeff_1}
\end{eqnarray}
For non-degenerate eigenstates $\Ket{n;r}$ and $\Ket{m;r}$, one can express the second matrix element on the right-hand side of (\ref{Eq:SE_coeff_1}) as
\begin{equation}
 \Braket{n;r|\partial_r|m;r} = 
 \begin{cases}
  \frac{\Braket{n;r|(\partial_r M(r))|m;r}}{q_m(r) - q_n(r)} & \textrm{  for } n \neq m\\ 
  \Braket{n;r|\partial_r|n;r} & \textrm{  for } n = m \textrm{.}
 \end{cases}
\label{Eq:diabatic_coupling_1}
\end{equation}
It is now interesting to notice that the representation of $M(r)$ in the eigenbasis of the static circular billiard (\ref{Eq:eigenfunctions_circular_billiard}),
$\Braket{\Phi_{n',m'}|M(r)|\Phi_{n,m}}$,
is not only a Hermitian, but a real symmetric matrix (cf. Eq.~(\ref{Eq:mathieu_elements}) in appendix \ref{App:matrix_elements}). 
We can therefore choose the expansion coefficients of the eigenstates of $M(r)$ in the eigenbasis of the static circular billiard, $\Braket{\Phi_{n',m'}|n;r}$, to be real.
It follows that also the expansion coefficients of $\partial_r\Ket{n;r}$, $\Braket{\Phi_{n',m'}|\partial_r|n;r}$, are real.
Thus,
\begin{equation}
 \Braket{n;r|\partial_r|n;r} = \sum_{n',m'} \Braket{n;r|\Phi_{n',m'}}\Braket{\Phi_{n',m'}|\partial_r|n;r}
\label{Eq:Berry_expansion_circular}
\end{equation}
is also real.
On the other hand, due to normalization of the eigenstates $\Ket{n;r}$, expression (\ref{Eq:Berry_expansion_circular}) has to be purely imaginary and is therefore identical zero.
Noting that
\begin{eqnarray}
 \partial_r M(r) &=& \frac{\partial^2}{\partial \eta^2} - \frac{1}{r^2}\frac{\partial^2}{\partial \xi^2} \nonumber\\
&=& \frac{1}{r} \left(r \frac{\partial^2}{\partial \eta^2} - \frac{1}{r}\frac{\partial^2}{\partial \xi^2} \right) \nonumber\\
&=& \frac{M(\dot \imath r)}{\dot \imath r}
\label{Eq:parameter_derivative_M}
\end{eqnarray}
further simplifies Eq.~(\ref{Eq:diabatic_coupling_1}).
From now on, we will restrict ourselves to periodic driving laws,  i.e. $a(t+\frac{2 \pi}{\omega}) = a(t)$ and $b(t+\frac{2 \pi}{\omega}) = b(t)$,
and cases where all populated instantaneous eigenstates are non-degenerate.
These conditions also include all cases that are discussed in \cite{Flrespop}.
We further introduce a rescaled dimensionless time $\tau := \frac{\omega}{2 \pi} t$ and finally put Eqs.~(\ref{Eq:diabatic_coupling_1}) and (\ref{Eq:parameter_derivative_M}) back into (\ref{Eq:SE_coeff_1}).
\begin{eqnarray}
 \dot \imath \partial_{\tau} c_n(\tau) &=& \sum_m c_m(\tau) e^{-\frac{2 \pi \dot \imath}{\hbar \omega} (\phi_m(\tau)-\phi_n(\tau))} \frac{\omega}{2 \pi \hbar} \Braket{n;r(\tau)|H_F(\tau)|m;r(\tau)} \nonumber\\
&+& \sum_{m \neq n} c_m(\tau) e^{-\frac{2 \pi \dot \imath}{\hbar \omega} (\phi_m(\tau)-\phi_n(\tau))} \frac{\dot r(\tau)}{r(\tau)} \frac{\Braket{n;r(\tau)|M(\dot \imath r(\tau))|m;r(\tau)}}{q_n(r(\tau)) - q_m(r(\tau))} \textrm{.}
\label{Eq:SE_coeff_2}
\end{eqnarray}
We point out that the modulus of the first term in (\ref{Eq:SE_coeff_2}) depends linearly on the driving frequency $\omega$ while the modulus of the second term is independent of $\omega$.
We therefore expect the first term to be dominating for large driving frequencies, while the second one should be dominant for small driving frequencies and should especially couple 
neighboring instantaneous eigenstates due to the denominator $q_n(r(\tau)) - q_m(r(\tau))$.

Obviously, due to periodic driving, all terms on the right-hand side of (\ref{Eq:SE_coeff_2}) but the coefficients $c_m(\tau)$ and the phase factors 
$\exp(-2 \pi \dot \imath (\phi_m(\tau)-\phi_n(\tau))/\hbar \omega)$ are one-periodic functions in $\tau$.
It is therefore possible to represent them by discrete Fourier transforms.
Before we do so, we split $\phi_m(\tau)-\phi_n(\tau)$ into a non-periodic part $\hbar \nu_{mn} \tau := (\phi_m(1)-\phi_n(1)) \tau $ and a one-periodic part $\hbar \Delta\nu_{mn}(\tau)$:
\begin{equation}
 \phi_m(\tau)-\phi_n(\tau) = \hbar \nu_{mn} \tau + \hbar \Delta\nu_{mn}(\tau) \textrm{.}
\label{Eq:split_phi}
\end{equation}
We then combine the one-periodic phase factor $\exp(-2 \pi \dot \imath \Delta\nu_{mn}(\tau)/\omega)$ with the other one-periodic terms on the right-hand side of (\ref{Eq:SE_coeff_2})
and Fourier transform the results:
\begin{eqnarray}
 \omega \sum_{l = -\infty}^{l = \infty} F_l^{nm} e^{-2 \pi \dot \imath l \tau} &=& e^{-\frac{2 \pi \dot \imath}{\omega} \Delta\nu_{mn}(\tau)} \frac{\omega}{2 \pi \hbar} \Braket{n;r(\tau)|H_F(\tau)|m;r(\tau)}
\label{Eq:FT_pert_coupling} \\
 \sum_{l = -\infty}^{l = \infty} D_l^{nm} e^{-2 \pi \dot \imath l \tau} &=& 
\begin{cases}
  e^{-\frac{2 \pi \dot \imath}{\omega} \Delta\nu_{mn}(\tau)} \frac{\dot r(\tau)}{r(\tau)} \frac{\Braket{n;r(\tau)|M(\dot \imath r(\tau))|m;r(\tau)}}{q_n(r(\tau)) - q_m(r(\tau))} & \textrm{  for } n \neq m\\ 
  0 & \textrm{  for } n = m \textrm{.}
\end{cases}
\label{Eq:FT_diab_coupling}
\end{eqnarray}

Before we put (\ref{Eq:FT_pert_coupling}) and (\ref{Eq:FT_diab_coupling}) back into the SE (\ref{Eq:SE_coeff_2}), it is useful to perform a unitary transformation,
$c_n(\tau) = \exp(-\dot \imath \omega F_{0}^{nn} \tau) b_n(\tau)$.
Note that we do not include $D_0^{nn}$ in the unitary transformation as it is zero by definition (\ref{Eq:FT_diab_coupling}) and that $F_{0}^{nn}$ is completely independent of $\omega$.
This unitary transformation, together with Eqs.~(\ref{Eq:FT_pert_coupling}) and (\ref{Eq:FT_diab_coupling}), leads via (\ref{Eq:SE_coeff_2}) to a SE for the coefficients $b_n$: 
\begin{equation}
 \dot \imath \dot b_n(\tau) = \sum_{\substack{m,l \\ m \neq n \textrm{ for } l = 0}} e^{2 \pi \dot \imath \theta_{l}^{nm} \tau} (\omega F_l^{nm} + D_l^{nm}) b_m(\tau)  \textrm{,}
\label{Eq:FT_SE_1}
\end{equation}
where we have defined the abbreviation
\begin{equation}
 \theta_{l}^{nm} := \frac{\nu_{nm}}{\omega} + \frac{\omega}{2 \pi} (F_{0}^{nn} - F_{0}^{mm}) - l \textrm{.}
\label{Eq:theta_def}
\end{equation}
Note that the solution of Eq.~(\ref{Eq:FT_SE_1}) determines the complete physics of periodically driven elliptical quantum billiards.

\subsection{Perturbative analysis}
\label{ch:perturbative_analysis}

We will now use time-dependent perturbation theory (TDPT) to find an approximate solution of Eq.~(\ref{Eq:FT_SE_1}) in first order.
To do so, we formally affix a parameter $\lambda$ to $F_l^{nm}$ and $D_l^{nm}$ to keep track of the order of perturbation and will set $\lambda$ to 1 at the end of our calculation:
$F_l^{nm} = \lambda \cdot F_l^{nm}$, $D_l^{nm} = \lambda \cdot D_l^{nm}$.
An expansion of $b_n(\tau)$ in $\lambda$ gives: $b_n(\tau) = \sum_{p = 0}^{\infty} \lambda^{p} b_n^{(p)}(\tau)$.
As $\lambda$ should track the order of perturbation, it is natural to choose the initial values of $b_n^{(p)}$ as $b_n^{(p)}(0) = \delta_{p,0} \cdot b_n(0)$.
Inserting this ansatz into (\ref{Eq:FT_SE_1}) and equating equal powers of $\lambda$ yields up to first order:
\begin{eqnarray}
 \dot \imath \dot b_n^{(0)} &=& 0 \Rightarrow b_n^{(0)} = \textrm{const.} = b_n(0) \\
 \nonumber\\
 \dot \imath \dot b_n^{(1)} &=& \sum_{\substack{m,l \\ m \neq n \textrm{ for } l = 0}} e^{2 \pi \dot \imath \theta_{l}^{nm} \tau} (\omega F_l^{nm} + D_l^{nm}) b_m(0) \\
 \nonumber\\
 \Rightarrow b_n^{(1)}(\tau) &=& \sum_{\substack{m,l \\ m \neq n \textrm{ for } l = 0}} \frac{e^{2 \pi \dot \imath \theta_{l}^{nm} \tau} - 1}{2 \pi \dot \imath \theta_{l}^{nm}} (\omega F_l^{nm} + D_l^{nm}) b_m(0) \label{Eq:first_order_coeff} \ \textrm{.}
\end{eqnarray}

\subsubsection{Population transfer probability}

We are now able to calculate the population transfer probability between two instantaneous eigenstates which will lead to a systematical understanding of resonant population transfer as it was observed
in \cite{Flrespop}.
For this purpose we assume that the wave function $\Ket{\Lambda}$ was initially in the (undriven) eigenstate
$\Ket{k;r}$ and then calculate the evolution of the population of the eigenstate $\Ket{n;r}$ ($n \neq k$).
Population transfer in first order gives:
\begin{equation}
 p_{nk}^{1}(\tau) := |b_n^{(1)}(\tau)|^2 = \sum_{l,l'} \frac{e^{2 \pi \dot \imath \theta_{l}^{nk} \tau} - 1}{2 \pi \theta_{l}^{nk}} \cdot \frac{e^{- 2 \pi \dot \imath \theta_{l'}^{nk} \tau} - 1}{2 \pi \theta_{l'}^{nk}} (\omega F_l^{nk} + D_l^{nk})(\omega F_{l'}^{*nk} + D_{l'}^{*nk}) \textrm{.}
\label{Eq:pnk_1}
\end{equation}
We would like to calculate a population transition rate per unit time from (\ref{Eq:pnk_1}) which is defined as $\Gamma_{nk}^{1} := \lim_{\tau \rightarrow \infty} p_{nk}^{1}(\tau)/\tau$.
Note that $(e^{2 \pi \dot \imath \theta_{l}^{nk} \tau} - 1)/2 \pi \theta_{l}^{nk} = \dot \imath e^{\pi \dot \imath \theta_{l}^{nk} \tau} \sin(\pi \theta_{l}^{nk} \tau)/(\pi \theta_{l}^{nk})$
grows linearly with $\tau$ for $\theta_{l}^{nk} = 0$ while it oscillates periodically with an amplitude $1/\pi \theta_{l}^{nk}$ (which is independent of $\tau$) for  $\theta_{l}^{nk} \neq 0$.
Due to $\theta_{l}^{nk} - \theta_{l'}^{nk} = l' -l$, we can therefore neglect all terms in (\ref{Eq:pnk_1}) with $l \neq l'$ for $\tau$ being sufficiently large. 

\begin{eqnarray}
 \Gamma^{nk}_{1} &:=& \lim_{\tau \rightarrow \infty} \frac{p_{nk}^{1}(\tau)}{\tau} \nonumber\\
 &=& \lim_{\tau \rightarrow \infty} \sum_l \frac{\sin^2 \pi \theta_{l}^{nk} \tau}{\tau (\pi \theta_{l}^{nk})^2} \left|\omega F_l^{nk} + D_l^{nk}\right|^2 \nonumber\\
 &=& \sum_l \delta(\theta_{l}^{nk}) \left|\omega F_l^{nk} + D_l^{nk}\right|^2 \textrm{.}
 \label{Eq:transition_rate_calc}
\end{eqnarray}
By applying appropriate transformations, we have handled the time-dependent boundary conditions of the billiard by introduction of a time-dependent external potential.
This enabled us to derive Eq.~(\ref{Eq:transition_rate_calc}) which is a Fermi's Golden Rule \cite{goldenRule} for driven elliptical quantum billiards. 
It states that efficient population transfer in first order between the instantaneous eigenstates $\Ket{k;r}$ and $\Ket{n;r}$ is only possible for $\theta_{l}^{nk}=0$.
We can now use (\ref{Eq:theta_def}) to calculate corresponding resonance frequencies,
\begin{equation}
 \omega_{res}^{nk,l} = \frac{l \pm \sqrt{l^2 - 4 \nu_{nk} \delta F_0^{nk}}}{2 \delta F_0^{nk}} \textrm{,}
\label{Eq:exact_resonance cond}
\end{equation}
where $2 \pi \cdot \delta F_0^{nk} := F_0^{nn} - F_0^{kk}$ has been defined.
Numerical experience shows that $\delta F_0^{nk}$ is usually a very small quantity.
The ``$+$''-term in (\ref{Eq:exact_resonance cond}) thus corresponds to a very large resonance frequency. 
Restricting ourselves to not too strongly driven billiards, we will neglect this term from now on. 
If we develop  the ``$-$''-term in (\ref{Eq:exact_resonance cond}) about $\delta F_0^{nk} \approx 0$ and use the definition of $\nu_{nk}$ in Eq.~(\ref{Eq:split_phi}) above, we find:
\begin{equation}
 l \cdot \omega_{res}^{nk,l} = \nu_{nk} = \int_0^1 E_n(\tau') - E_k(\tau') d\tau'
\label{Eq:appr_resonance_cond}
\end{equation}
Thus, only when the one-period average difference of two instantaneous energy eigenvalues matches an integer multiple of the driving frequency, resonant population transfer
between the corresponding instantaneous eigenstates can occur.
This is precisely the empirically found criterion in \cite{Flrespop} and has herewith a theoretical basis.
The result justifies to call the Fourier summation index $l$ ``photon process order'' of a population transfer in analogy to the interaction of light and matter.

\subsubsection{Applicability of first order TDPT}

Not all predicted resonance frequencies (\ref{Eq:exact_resonance cond}) are of equal importance with respect to their experimental observation and we will now derive a criterion to discriminate them. 
In the resonant case $\theta_{l}^{nk}=\theta_{l'}^{nk}=0$, Eq.~(\ref{Eq:pnk_1}) reduces to:
\begin{equation}
 p_{nk}^{1} = \tau^2 \sum_l \left|\omega F_l^{nk} + D_l^{nk}\right|^2
\label{Eq:p_nk_in_res}
\end{equation}
The reader is reminded that (\ref{Eq:p_nk_in_res}) only holds for $n \neq k$, while for $n = k$, $p_{kk}^{1}=0$ holds as $\theta_{l}^{kk}=0$ implies $l=0$ and this case just had been excluded 
from the summation in Eq.~(\ref{Eq:first_order_coeff}).
Consequently, the instantaneous eigenstate $\Ket{k;r}$ gets exclusively depopulated in first order TDPT.
We can therefore calculate the time $\tau_{\textrm{int}}$ at which the population $p_k$ of the instantaneous eigenstate $\Ket{k;r}$ gets negative and therefore unphysical:
\begin{eqnarray}
 p_k(\tau_{\textrm{int}}) = 1 - \sum_{\theta_{l}^{nk} \equiv 0} p_{nk}^{1}(\tau_{\textrm{int}}) \stackrel{!}{=} 0 \label{Eq:p_k} \\
\Rightarrow \tau_{\textrm{int}} = \frac{1}{\sqrt{\sum_{\theta_{l}^{nk} \equiv 0} \left|\omega F_l^{nk} + D_l^{nk}\right|^2}} \textrm{.}
\label{Eq:tau_int}
\end{eqnarray}
The summation index $\theta_{l}^{nk} \equiv 0$ in (\ref{Eq:p_k}) and (\ref{Eq:tau_int}) means that it should only be summed over states $n$ and photon process orders $l$ that satisfy the 
resonance condition $\theta_{l}^{nk} \equiv 0$.
This means in all practical examples that the sum only consists of a single term.

$\tau_{\textrm{int}}$ is a measure for how fast a population probability transfer takes place.
It is, thus, reasonable that we will not be able to fully resolve resonances that correspond to an interaction time $\tau_{\textrm{int}}$ 
that is much larger than the actual runtime $\tau_{\textrm{run}}$ of a possible experiment.
In this case, population transfer will have been stopped before the maximal theoretically 
possible amount of population probability will have been transferred from one instantaneous eigenstate to the other and
our ability to resolve a resonance in corresponding observations is diminished.

On the other hand, we understand that the transition rate (\ref{Eq:transition_rate_calc}) has been calculated in the limit $\tau \to \infty$ and the included $\delta$-function 
is the result of a convergence process.
In order to have the system meet the predictions of first order TDPT, $\tau_{\textrm{int}}$ should be large enough such that a delta function $\delta(\theta)$ is a good approximation of 
$\sin^2 \pi \theta \tau/\tau (\pi \theta)^2$ as it appears in the derivation of (\ref{Eq:transition_rate_calc}). 
Obviously, such criterion depends on the density of the $\theta_{l}^{nk}$ about $\theta = 0$.
We therefore define a lower threshold
\begin{equation}
 \tau_{\textrm{low}} := \max_{|\theta_{l}^{nk}| \neq 0} \frac{1}{|\theta_{l}^{nk}|}
\label{Eq:def_tau_low}
\end{equation}
where only $\theta_{l}^{nk}$ should be considered in (\ref{Eq:def_tau_low}) whose corresponding resonant probability transitions (i.e. for the case $\theta_{l}^{nk}=0$) have interaction times 
of the order of magnitude of $\tau_{\textrm{run}}$ such that they are relevant for the experiment.
In summary, we expect predicted resonances to be fully resolved if
\begin{equation}
 \tau_{\textrm{low}} \ll \tau_{\textrm{int}} < \tau_{\textrm{run}}
\label{Eq:res_criterion}
\end{equation}
holds.
The discussion of concrete driving laws in chapter \ref{ch:numerical_results} shows that this criterion is in excellent agreement with our numerical simulations.

\subsubsection{Rotating wave approximation}

Interestingly, (\ref{Eq:res_criterion}) justifies a rotating wave approximation in (\ref{Eq:FT_SE_1}) \cite{vogel2006quantum}.
This allows us to calculate approximate population dynamics of the system that, in contrast to (\ref{Eq:pnk_1}), conserve the total population probability.

For simplicity, we will assume that there is only one $\theta_{l}^{nk}$ close to zero.
A rotating wave approximation simply sets all other terms in (\ref{Eq:FT_SE_1}) that do not contain this $\theta_{l}^{nk}$ to zero as they are comparatively fast oscillating,
such that we are left with with an effective two-level system:

\begin{eqnarray}
 \dot \imath \dot b_n(\tau) &=& e^{2 \pi \dot \imath \theta_{l}^{nk} \tau}(\omega F_l^{nk} + D_l^{nk}) b_k(\tau) \nonumber\\
 \dot \imath \dot b_k(\tau) &=& e^{-2 \pi \dot \imath \theta_{l}^{nk} \tau} (\omega F_l^{kn} + D_l^{kn}) b_n(\tau)  \ \textrm{.} 
 \label{Eq:rabi_eq}
\end{eqnarray}

The behavior of such a system is very well understood. A discussion in terms of Bloch equations is, for instance, given in \cite{mandel1995optical}. 
Eqs.~(\ref{Eq:rabi_eq}) especially explain why the population dynamics in \cite{Flrespop} are reminiscent of Rabi oscillations.
The effective Rabi-frequency $\Omega_{\textrm{eff}}$ can be calculated (cf. \cite{mandel1995optical}) to be
\begin{equation}
 \Omega_{\textrm{eff}} = \sqrt{\left(2 \pi \theta_{l}^{nk}\right)^2 + 4 |\omega F_l^{nk} + D_l^{nk}|^2} = \sqrt{\left(2 \pi \theta_{l}^{nk}\right)^2 + 4 \frac{1}{\tau_{\textrm{int}}^2}}
\end{equation}
which yields a beating period $T_B$ of the population dynamics
\begin{equation}
 T_B := \frac{2 \pi}{\Omega} = \frac{\pi \tau_{\textrm{int}}}{\sqrt{1 + \left(\pi \theta_{l}^{nk} \tau_{\textrm{int}}\right)^2}} \textrm{.}
\label{Eq:Beating_period}
\end{equation}
In summary, if we assume the system to have initially been in state $k$, the population dynamics of state $n$ is given by:
\begin{equation}
 p_n(\tau) = \frac{\sin^2\left(\frac{\pi \tau}{T_B}\right)}{1 + (\pi \theta_{l}^{nk} \tau_{\textrm{int}})^2} \textrm{.}
\label{Eq:rabi_population} 
\end{equation}

\section{Numerical results and discussion}
\label{ch:numerical_results}

In this chapter we will present full numerical simulations of driven elliptical billiards and analyze the results with the developed perturbation theory of chapter \ref{ch:perturbative_analysis}.
Details to the numerical calculation of the predicted quantities can be found in appendix \ref{App:matrix_elements}.
All numerical calculations have been run for $\tau_{\textrm{run}} = 100$ periods of driving and $\hbar$ and $\mu$ have, w.l.o.g., been set to 1.
We will always drive the semi-axis $a(t)$ harmonically, i.e.
\begin{equation}
 a(t) = a_0 + A \sin(\omega t) \textrm{,}
\label{Eq:a(t)}
\end{equation}
and adjust $b(t)$ such that the billiard is driven in different ways as will be specified later.
To be able to compare the different driving laws, we have chosen to keep the following parameters fixed: 
\begin{equation}
 a_0 := a(t=0) = 1\textrm{, } A = 0.1 \textrm{ and } b_0 := b(t=0) = \sqrt{0.51} \textrm{.}
\label{Eq:parameter_choices}
\end{equation}
These parameters are the same as in \cite{Flrespop}.

The energy $E(\tau)$ will be a key observable for the analysis of the billiard dynamics.
It is calculated from the expectation value of the Hamiltonian $H = -\frac{\hbar^2}{2 \mu} \Delta$ as it appears in Eq.~(\ref{Eq:SE1}):
\begin{equation}
 E(\tau) = \Braket{\Psi(\tau)|-\frac{\hbar^2}{2 \mu} \Delta|\Psi(\tau)} \textrm{.}
\end{equation}
If we apply again the coordinate transformation (\ref{Eq:coordtrafo}) and the unitary transformation $U$ (\ref{Eq:U}), the energy reads
\begin{equation}
 E(\tau) = \Braket{\Lambda(\tau)|U^{\dagger}(\tau)H_M(\tau)U(\tau)|\Lambda(\tau)} \textrm{,}
\end{equation}
where $H_M(\tau)$ is given by (\ref{Eq:Mathieu_Hamiltonian}).
We can therefore calculate $E(\tau)$ by determining the population $p_n(\tau)$ of the eigenstates of $U^{\dagger}(\tau)H_M(\tau)U(\tau)$ in $\Ket{\Lambda(\tau)}$,
weighting these populations with the respective eigenvalues $E_n(\tau)$ of $U^{\dagger}(\tau)H_M(\tau)U(\tau)$ and sum the results up:
\begin{equation}
 E(\tau) = \sum_n E_n(\tau) p_n(\tau) \textrm{.}
\label{Eq:energy_by_pop}
\end{equation}
Note that, due to $U$ being a unitary transformation, the energy eigenvalues $E_n(\tau)$ of $U^{\dagger}(\tau)H_M(\tau)U(\tau)$ are actually identical to the instantaneous eigenvalues $E_n(\tau)$ of $H_M(\tau)$.

From now on, we refer to the instantaneous eigenstates of $U^{\dagger}(\tau)H_M(\tau)U(\tau)$ as energy eigenstates.
We understand in particular that the instantaneous eigenstates $\Ket{n;r(\tau)}$ of $H_M(\tau)$ are in general not identical to the energy eigenstates, 
but unitarily transformed energy eigenstates, given by $U^{\dagger}\Ket{n;r(\tau)}$.
Note that $U$ is also invariant upon sign-change of $\eta$ and $\xi$, such that an instantaneous eigenstate $\Ket{n;r(\tau)}$ and its corresponding energy eigenstate $U^{\dagger}\Ket{n;r(\tau)}$
have the same $\eta$- and $\xi$- parity.

We will initialize the system in the fourth energy eigenstate (at $\tau=0$) and calculate the populations $p_n(\tau)$ upon driving.
Note that in \cite{Flrespop} instantaneous eigenstates $\Ket{n;r(\tau)}$ were used as initial states and for population analyses.

Interestingly, we find that the overlap $|\Braket{n;r(\tau)|U(\tau)|n;r(\tau)}|^2$ of all relevant instantaneous eigenstates $\Ket{n;r(\tau)}$ with their respective energy eigenstates is greater than 94.5\%
for the later analyzed parameter regimes in chapter \ref{ch:numerical_results}.
We thus expect that the energy eigenstates are similar to the instantaneous eigenstates and also show similar population dynamics.
Consequently, we will from now on disregard the differences between $\Ket{n;r(\tau)}$ and the energy eigenstates $U^{\dagger}\Ket{n;r(\tau)}$ when analyzing the billiard dynamics perturbatively and 
will subsequently refer to $\Ket{n;r(\tau)}$ simply as energy eigenstates.
This approximation enables us to predict the seemingly complicated population dynamics of the energy eigenstates by our perturbation theory for instantaneous eigenstates.
Although the population dynamics of the energy eigenstates are not expected to be qualitatively different, some predictions may be compromised quantitatively.
For instance, due to the different actions of $U^{\dagger}(\tau)$ on different $\Ket{n;r(\tau)}$, shifts of the resonance frequencies (\ref{Eq:exact_resonance cond}) are to be expected.
However, as $U$ (\ref{Eq:U}) approaches unity for vanishing $\omega$, these shifts will rather be observed for larger resonance frequencies.
We will also find that the resonance shifts become more negligible for higher order photon processes.

In the approximation of instantaneous eigenstates $\Ket{n;r(\tau)}$ being energy eigenstates, a transition to a higher excited state increases the energy $E(\tau)$ (\ref{Eq:energy_by_pop})
while a transition to a lower excited state decreases it.
We can therefore determine if a population transition occurs at a certain driving frequency $\omega$ upon simulation time $\tau_{\textrm{run}}$ by recording the maximal and minimal energy of the billiard
in dependence of $\omega$.

\subsection{Axes-ratio-preserving driving law}
\label{ch:axes_ratio}

In the following, we provide numerical solutions for various driving laws and analyze them with the developed perturbation theory of chapter \ref{ch:perturbative_analysis}.
A simple but illustrative driving law is the so-called axes-ratio-preserving driving law which merely rescales the billiard by varying its volume $V(\tau)$ while
keeping the ratio of the semi-axes $r(\tau)$ constant for all times upon driving.
As $a(t)$ is given by Eq.~(\ref{Eq:a(t)}), we find $ b(t) = r_0 a(t)$ and choose $r_0 = \sqrt{0.51}$ to satisfy Eq.~(\ref{Eq:parameter_choices}). 

\begin{figure}
 \begin{center}
 \includegraphics[width=0.49\textwidth]{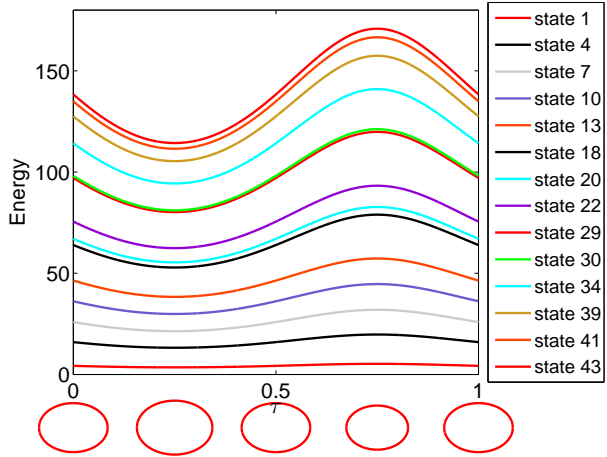}
 \end{center}
 \caption{Energy eigenvalues $E_n(\tau)$ of eigenstates $\Ket{n;r(\tau)}$ with even $\eta$- and $\xi$- parity eigenvalues for the axes-ratio-preserving driving law. 
The instantaneous shape of the ellipse at five different values of $\tau$ is drawn below the energy eigenvalue curves.
Parameters: $a_0 = 1$, $b_0 = \sqrt{0.51}$, $A = 0.1$.}
\label{Fig:ratio_eigenvalues}
\end{figure}

The axes-ratio-preserving driving law has the nice property that the instantaneous eigenstates $\Ket{n;r(\tau)}$ become time-independent due to fixed $r(\tau) = r_0$,
while the eigenvalues $E_n(\tau)$ stay time-dependent as can be seen from Eq.~(\ref{Eq:Mathieu_Ham_diagonal_form}).
Their variation is solely given by the global prefactor $1/V(\tau)$ which particularly prevents crossings of energy eigenvalues.
Fig.~\ref{Fig:ratio_eigenvalues} shows a sample of eigenvalue curves for one period of driving.
The fourth energy eigenstate has even $\eta$- and $\xi$- parity. 
Thus, only energy eigenstates in the corresponding sub-Hilbert space couple to the chosen initial state.

\begin{figure}
 \begin{center}
 \includegraphics[width=0.49\textwidth]{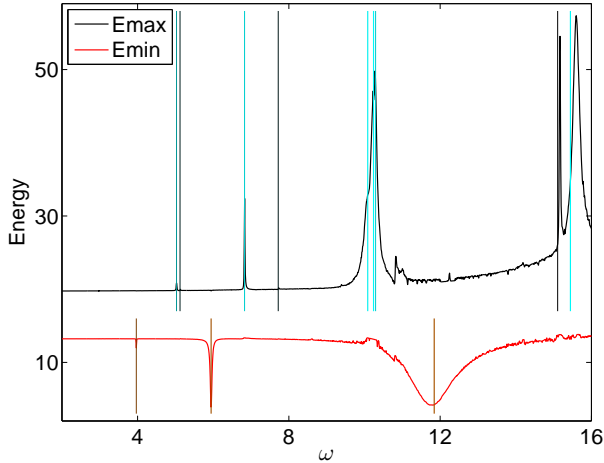}
 \end{center}
\caption{Dependence of the maximal and minimal energy on the driving frequency $\omega$ for the axes-ratio-preserving driving law.
The vertical lines show all predicted resonance frequencies (\ref{Eq:exact_resonance cond}) with an interaction time $\tau_{\textrm{int}}$ of less than 2000.
The darker the lines are, the longer is the corresponding interaction time.
Numerical values can be taken from Tab. \ref{Tab:ratio_resonances} in appendix \ref{App:theoretical_predictions}.
Parameters like in Fig.~\ref{Fig:ratio_eigenvalues}.}
\label{Fig:ratio_resonances}
\end{figure}

In Fig.~\ref{Fig:ratio_resonances} the dependence of the maximal and minimal energy that has been reached upon driving 
as a solution of the TDSE in Eq.~(\ref{Eq:SE1})
is plotted depending of the driving frequency $\omega$.
We clearly see sharp peaks and dips at certain driving frequencies. 
The vertical lines represent our predictions of resonance frequencies according to Eq.~(\ref{Eq:exact_resonance cond}).
Note that the observed resonances deviate slightly from the predicted ones, especially for larger driving frequencies.
This is due to the unitary transformation $U$ (\ref{Eq:U}) that has been neglected in our considerations, i.e. we apply perturbation theory 
only to the instantaneous eigenstate $\Ket{n;r(\tau)}$ that is most populated in the energy eigenstate $U^{\dagger}(\tau)\Ket{n;r(\tau)}$.
Beside this anticipated small deviation, we find a very good agreement of the numerical calculations with our predictions.
All resonances with a comparatively small interaction time $\tau_\textrm{int}$ have been resolved, while resonances with very large interaction times could not be observed.
Naturally, for interaction times that are longer than (half) the runtime of an experiment $\tau_{\textrm{run}}$, a full population transition from the initial state to some other energy eigenstate
can not happen according to Eq.~(\ref{Eq:rabi_population}). 
This is the reason why some peaks in Fig.~\ref{Fig:ratio_resonances} that correspond to transitions to the same energy eigenstate possess different heights.
It is interesting to note that, although the runtime $\tau_{\textrm{run}}$ of our numerical simulations was only 100 periods of driving, resonances that correspond to an interaction time of up to 2000 
periods of driving could still be partly resolved in Fig.~\ref{Fig:ratio_resonances}.
Tab. \ref{Tab:ratio_resonances} in appendix \ref{App:theoretical_predictions} provides numerical values for all predicted resonance frequencies between 0 and 16 that have an interaction time of less than 2000.
It also shows that the lower threshold $\tau_{\textrm{low}}$ is always much smaller than the corresponding interaction time $\tau_\textrm{int}$, such that the first part of Eq.~(\ref{Eq:res_criterion})
is fulfilled and TDPT of first order is applicable.

\begin{figure}
 \begin{center}
 \includegraphics[width=0.49\textwidth]{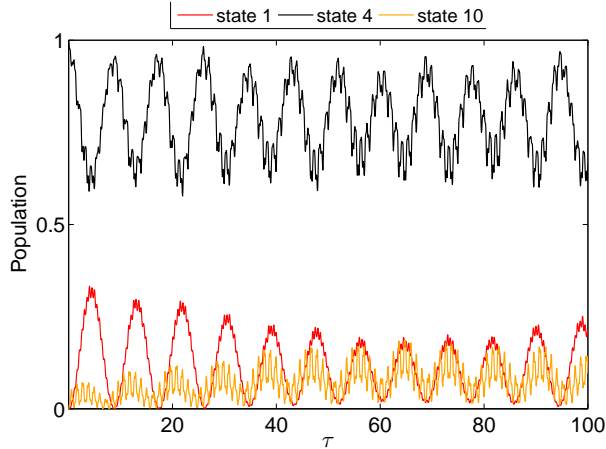}
 \end{center}
\caption{Population dynamics $p_n(\tau)$ for the axes-ratio-preserving driving law at $\omega = 10.81$.
Parameters like in Fig.~\ref{Fig:ratio_eigenvalues}.}
\label{Fig:ratio_pop_1}
\end{figure}

One might wonder about a structure of several small peaks, especially for frequencies $\omega \gtrsim 10.5$.
We assume that these smaller, not predicted peaks correspond to transitions of second order where population is first transferred to one excited state and then from this state again transferred 
to yet another energy eigenstate.
This is, for instance, supported by a population analysis in Fig.~\ref{Fig:ratio_pop_1} for the small peak at $\omega = 10.81$.
We see that the mean (or envelope behaviour) population of the energetical ground state decreases while the population amplitude of the tenth energy eigenstate increases.
This may be interpreted as an interaction of these two states that consecutively leads to a transfer of population that was initially transferred to the ground state and is then pushed to the tenth energy eigenstate.
Such a process is not included in the time-dependent perturbation theory of first-order in chapter \ref{ch:perturbative_analysis} and the dynamics visualized in Fig.~\ref{Fig:ratio_pop_1} are a precursor 
to the breakdown of this simple theory for higher driving frequencies where indirect transitions become more and more important.

It is also interesting that the resonance at $\omega \approx 15.17$ that corresponds to a 4-photon transition from the initial state to the 22nd energy eigenstate is so well resolved 
although the interaction time of this resonance is much larger than the interaction time of several resonances that are much worse resolved.
The reason for this is that the 22nd energy eigenstate has a much higher energy eigenvalue than, for instance, the seventh energy eigenstate.
Thus, the energy is significantly increased for already a small amount of transferred population probability from the initial state to the 22nd energy eigenstate.

\begin{figure}
 \begin{center}
 \includegraphics[width=0.49\textwidth]{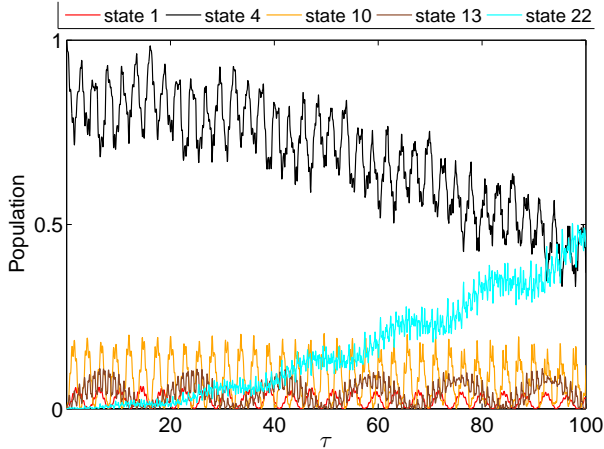}
 \end{center}
\caption{Population dynamics $p_n(\tau)$ for the axes-ratio-preserving driving law at $\omega = 15.17$.
Parameters like in Fig.~\ref{Fig:ratio_eigenvalues}.}
\label{Fig:ratio_pop_2}
\end{figure}

\subsection{Other driving laws}
\label{ch:other}

\begin{figure}
 \begin{center}
 \includegraphics[width=0.49\textwidth]{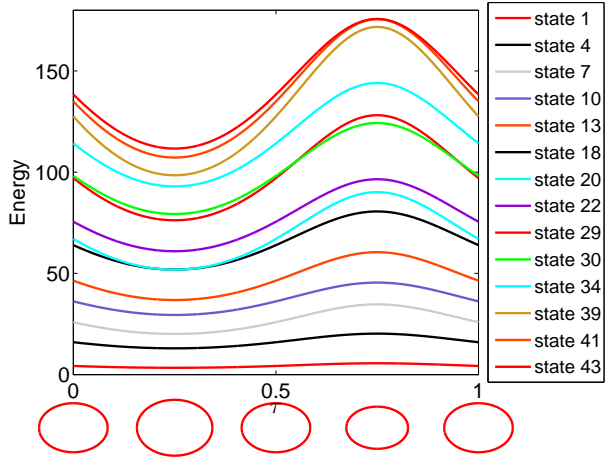}
 \end{center}
 \caption{Energy eigenvalues $E_n(\tau)$ of eigenstates $\Ket{n;r(\tau)}$ with even $\eta$- and $\xi$- parity for the breathing driving law.
Parameters like in Fig.~\ref{Fig:ratio_eigenvalues}.}
\label{Fig:breathing_eigenvalues}
\end{figure}

\begin{figure}
 \begin{center}
 \includegraphics[width=0.49\textwidth]{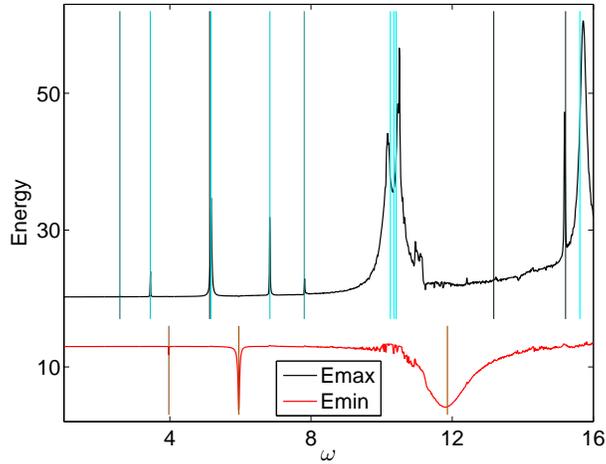}
 \end{center}
\caption{Analogue of Fig.~\ref{Fig:ratio_resonances} for the breathing driving law.
Numerical values can be taken from Tab. \ref{Tab:breathing_resonances} in appendix \ref{App:theoretical_predictions}.
Parameters like in Fig.~\ref{Fig:ratio_eigenvalues}.}
\label{Fig:breathing_resonances}
\end{figure}

We have seen in the last chapter that the predictions of TDPT work very well for the axes-ratio-preserving driving law.
To demonstrate the general applicability of the perturbation theory of chapter \ref{ch:perturbative_analysis}, we will analyze two further driving laws.
The so-called breathing driving law $b(t) = a(t) - a_0 + b_0$, where $a(t)$ is again given by Eq.~(\ref{Eq:a(t)}), was already discussed in \cite{Flrespop}.
Fig.~\ref{Fig:breathing_eigenvalues} shows the eigenvalues of energy eigenstates with even $\eta$- and $\xi$- parity for one period of driving.
We see that crossings of energy eigenvalues are, in contrast to Fig.~\ref{Fig:ratio_eigenvalues}, now possible as $r(t)$ is no longer kept constant.
The eigenvalues of the lowest excited states are, though, very similar to the ones in Fig.~\ref{Fig:ratio_eigenvalues} and, consequently, 
the resonances in Fig.~\ref{Fig:breathing_resonances} resemble the ones in Fig.~\ref{Fig:ratio_resonances}.
In Fig.~\ref{Fig:breathing_resonances} more resonances can be resolved due to a sufficiently small interaction time $\tau_{\textrm{int}}$.
This observation can be understood as follows: 
While for the axes-ratio-preserving driving law the transition matrix $D_l^{nm}$ (\ref{Eq:FT_diab_coupling}) is identical to zero due to $\dot r(t) = 0$, this is not the case for the breathing driving law.
The additionally resolved resonances for the breathing law correspond, thus, to Landau-Zener transitions with $\dot r(t)$ being the Landau-Zener velocity \cite{zener}.
This role of $D_l^{nm}$ triggering Landau-Zener transitions will be even more pronounced for the next driving law that is presented.

\begin{figure}
 \begin{center}
 \includegraphics[width=0.49\textwidth]{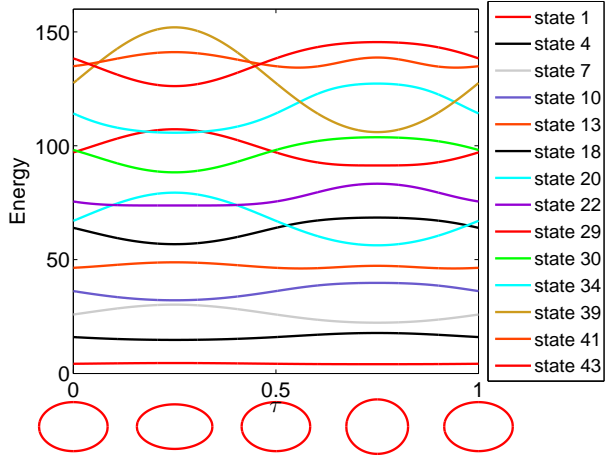}
 \end{center}
 \caption{Energy eigenvalues $E_n(\tau)$ of eigenstates $\Ket{n;r(\tau)}$ with even $\eta$- and $\xi$- parity eigenvalues for the volume-preserving driving law.
Parameters like in Fig.~\ref{Fig:ratio_eigenvalues}.}
\label{Fig:vol_eigenvalues}
\end{figure}

The so-called volume-preserving driving law is just the opposite of the axes-ratio-preserving driving law.
It keeps the volume $V(t)$ of the billiard fixed, while varying the ratio of the semi-axes $r(t)$.
Thus, $b(t)$ depends on $a(t)$ (\ref{Eq:a(t)}) as $b(t) = a_0 b_0/a(t)$.
Fig.~\ref{Fig:vol_eigenvalues} shows the corresponding energy eigenvalues.
We see that the fourth and seventh energy eigenvalue get close upon driving such that we expect that the transition matrix $D_l^{nm}$ (\ref{Eq:FT_diab_coupling}) couples these states strongly.
We point out that we can arbitrarily control how close these eigenvalues get upon driving by choosing $r(t)$ appropriately.
As $D_l^{nm}$ (\ref{Eq:FT_diab_coupling}) depends on $\omega$ only through the phase factor $e^{-\frac{2 \pi \dot \imath}{\omega} \Delta\nu_{mn}(\tau)}$,
we expect it to be slowly varying with $\omega$, thus setting an upper bound on the interaction time (\ref{Eq:tau_int}) of resonant population transitions between the fourth and seventh
energy eigenstate even for small $\omega$ and corresponding large photon process orders $l$.

\begin{figure}
 \begin{center}
 \includegraphics[width=0.49\textwidth]{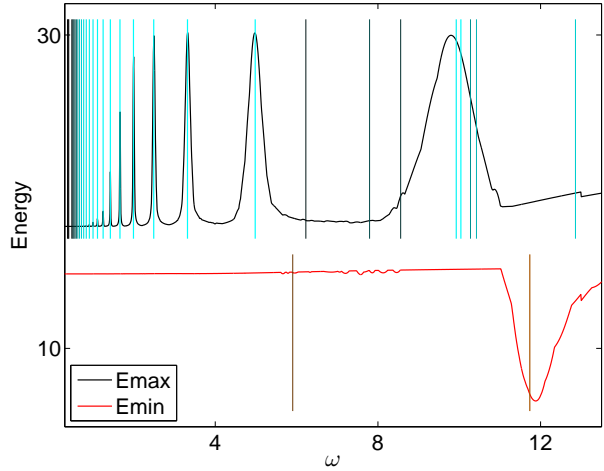}
 \end{center}
\caption{Analogue of Fig.~\ref{Fig:ratio_resonances} for the volume-preserving driving law.
Numerical values can be taken from Tab. \ref{Tab:vol_resonances} in appendix \ref{App:theoretical_predictions}.
Parameters like in Fig.~\ref{Fig:ratio_eigenvalues}.}
\label{Fig:vol_resonances}
\end{figure}

This expectation is fully confirmed by Fig.~\ref{Fig:vol_resonances}.
All resolved resonances with $\omega < 5.5$ correspond exclusively to transitions between the fourth and seventh energy eigenstate.
As $U$ (\ref{Eq:U}) gets close to identity for small driving frequencies $\omega$, the induced resonance shift is negligible.
One might wonder, why most of the resonances at small driving frequencies have numerically not been fully resolved although their interaction time is short enough.
In review of Eq.~(\ref{Eq:appr_resonance_cond}), we understand that a detuning in the driving frequency is also multiplied by the photon process order $l$.
Thus, one has to adjust the driving frequency very carefully to resolve a multiple photon resonance.
We assume that the frequency grid in Fig.~\ref{Fig:vol_resonances} is not fine enough to resolve all predicted resonances although it should in principle be possible.

\begin{figure}
 \begin{center}
 \includegraphics[width=0.49\textwidth]{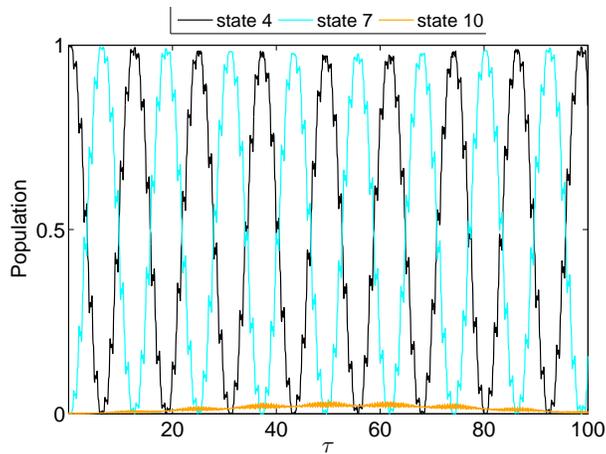}
 \end{center}
\caption{Population dynamics $p_n(\tau)$ for the volume-preserving driving law at $\omega = 3.32$.
Parameters like in Fig.~\ref{Fig:ratio_eigenvalues}.}
\label{Fig:vol_pop_2}
\end{figure}

Finally, population analyses close to the resonance frequencies confirm the Rabi-like behavior of the population dynamics as predicted by Eq.~(\ref{Eq:rabi_population}).
This can be especially well illustrated for resonances with small interaction times. 
As an example, Fig.~\ref{Fig:vol_pop_2} shows almost perfect Rabi-like population dynamics of the fourth and seventh energy eigenstate in the nearly resonant case of a three photon process at $\omega = 3.32$.
Comparison of the observed beating periods with the corresponding interaction times $\tau_{\textrm{int}}$ in Tab. \ref{Tab:vol_resonances} in appendix \ref{App:theoretical_predictions} gives,
even quantitatively, a very good agreement as predicted by Eq.~(\ref{Eq:Beating_period}).
We point out that as we can tune the strength of the transition matrix $D_l^{nm}$ by choosing how close the energy eigenvalues get upon driving, we can also tune the interaction time $\tau_{\textrm{int}}$
in the regime of weak driving where it is mainly determined by $D_l^{nm}$.
Hence, we can, in principle, also control the beating period of the present effective two-level Rabi-system.

\section{Brief Summary}
A time-dependent perturbative approach for elliptical quantum billiards with oscillating boundaries has been developed. 
As our major results we have obtained a Fermi Golden Rule, predicting the driving frequencies yielding resonant population transfer between instantaneous eigenstates
as observed in \cite{Flrespop} and a criterion allowing to decide which of these resonances are observable in a corresponding experiment of certain duration. 
Extensive numerical simulations have been performed for three different driving laws which are in excellent agreement with our predictions. 
Particularly for the volume preserving driving law, due to the change of the billiard geometry upon driving, Landau-Zener transitions have been analyzed to take place. 
Depending only weakly on the driving frequency, these transitions allow for resonant population transfer also for very weak driving. We have shown, that the billiard
dynamics can be reduced in this regime to an effective two-level system with in principle arbitrarily tunable oscillation period. 
\\Further interesting phenomena beyond the scope of our perturbative description can be expected in the 
numerically challanging regime of strong driving.

\section{Acknowledgements}
B.L. thanks the Landesexzellenzinitiative Hamburg "`Frontiers in Quantum Photon Science"',
which is funded by the Joachim Herz Stiftung, for financial support.

\clearpage

\appendix

\section{Matrix elements}
\label{App:matrix_elements}

Introducing the matrices
\begin{eqnarray}
 \hat{f}^1 &=& \sum_{n,m,n',m'} \Ket{\Phi_{n,m}}\delta_{m,m'} f^1_{nmn'}\Bra{\Phi_{n',m'}} \\
 \hat{f}^2 &=& \sum_{n,m,n',m'} \Ket{\Phi_{n,m}}\delta_{m,m'} f^2_{nmn'}\Bra{\Phi_{n',m'}} \\
 \hat{f}^3 &=& \sum_{n,m,n',m'} \Ket{\Phi_{n,m}}\delta_{(m-2),m'} f^3_{nmn'}\Bra{\Phi_{n',m'}} \\
 \hat{f}^4 &=& \sum_{n,m,n',m'} \Ket{\Phi_{n,m}}\delta_{(m-2),m'} f^4_{nmn'}\Bra{\Phi_{n',m'}} \\
 \hat{f}^5 &=& \sum_{n,m,n',m'} \Ket{\Phi_{n,m}}\delta_{(m+2),m'} f^5_{nmn'}\Bra{\Phi_{n',m'}} \\
 \hat{f}^6 &=& \sum_{n,m,n',m'} \Ket{\Phi_{n,m}}\delta_{(m+2),m'} f^6_{nmn'}\Bra{\Phi_{n',m'}} \textrm{,}
\end{eqnarray}
 with matrix elements
\begin{eqnarray}
 f^1_{nmn'} &=& \frac{-k^2_{m,n}}{4} \delta_{n,n'} \label{f1}\\
 f^2_{nmn'} &=& \frac{1}{2 J_{m+1}(k_{m,n}) J_{m+1}(k_{m,n'})} \int_0^1 J_m(k_{m,n}r) J_m(k_{m,n'}r) r^3 dr \label{f2}\\
 \nonumber\\
 f^3_{nmn'} &=& \frac{k^2_{m-2,n'}}{4 J_{m+1}(k_{m,n}) J_{m-1}(k_{m-2,n'})} \int_0^1 J_m(k_{m,n}r) J_m(k_{m-2,n'}r) r dr \label{f3}\\
 f^4_{nmn'} &=& \frac{1}{4 J_{m+1}(k_{m,n}) J_{m-1}(k_{m-2,n'})} \int_0^1 J_m(k_{m,n}r) J_{m-2}(k_{m-2,n'}r) r^3 dr \label{f4}\\
 \nonumber\\
 f^5_{nmn'} &=& \frac{k^2_{m+2,n'}}{4 J_{m+1}(k_{m,n}) J_{m+3}(k_{m+2,n'})} \int_0^1 J_m(k_{m,n}r) J_m(k_{m+2,n'}r) r dr \label{f5}\\
 f^6_{nmn'} &=& \frac{1}{4 J_{m+1}(k_{m,n}) J_{m+3}(k_{m+2,n'})} \int_0^1 J_m(k_{m,n}r) J_{m+2}(k_{m+2,n'}r) r^3 dr \label{f6} \textrm{,} 
\end{eqnarray}
where $J_m$ is again the cylindrical Bessel function of order $m$ and $k_{m,n}$ is its $n$-th root.
We have a convenient form of representing $H_M$ (\ref{Eq:Mathieu_Hamiltonian}), $M(r)$ (\ref{Eq:Mathieu_op}) and $H_F(\tau)$ (\ref{Eq:coupling_hamiltonian}) in the eigenbasis $\{\Ket{\Phi_{n,m}}\}_{n,m}$ 
(\ref{Eq:eigenfunctions_circular_billiard}) of the static circular billiard:
\begin{equation}
 H_M = g_1(\tau) \hat{f}^{1} + g_3(\tau) \left( \hat{f}^{3} + \hat{f}^{5} \right) \textrm{,}
\end{equation}
\begin{equation}
  M(r) \ = -\left(r + \frac{1}{r}\right) \hat{f}^{1} - \left(r - \frac{1}{r}\right) \left( \hat{f}^{3} + \hat{f}^{5} \right) \textrm{,}
\label{Eq:mathieu_elements}
\end{equation}
\begin{equation}
 H_F(\tau) \ = g_2(\tau) \hat{f}^{2} + g_4(\tau) \left(\hat{f}^{4} + \hat{f}^{6}\right) \textrm{.}
\label{H_pert_elements}
\end{equation}
Diagonalizing $M(r)$ yields the instantaneous eigenstates $\Ket{n;r}$ and their eigenvalues $q_n(r)$.
One could in principle calculate the energy eigenvalues $E_n(\tau) = \frac{\hbar^2}{\mu V(\tau)} q_n(r(\tau))$ from the $q_n(r)$,
but it turns out that diagonalizing $H_M$ directly increases the numerical precision of the energy eigenvalues.
Note that the time-dependent factors $g_i(\tau)$ as well as the matrix elements $f^i_{nmn'}$ are the same as in \cite{Flrespop}.
However, the matrix elements $f^i_{nmn'}$ have been reduced to a much simpler form, using orthonormality relations of the Bessel functions.
\begin{eqnarray}
 g_1(\tau) &=& -\frac{\hbar^2}{\mu}\left(\frac{1}{a(\tau)^2} + \frac{1}{b(\tau)^2}\right) \label{g1_t}\\
 g_2(\tau) &=& \mu \left(a(\tau)\ddot a(\tau) + b(\tau)\ddot b(\tau)\right) \label{g2_t}\\
 g_3(\tau) &=& -\frac{\hbar^2}{\mu}\left(\frac{1}{a(\tau)^2} - \frac{1}{b(\tau)^2}\right) \label{g3_t}\\
 g_4(\tau) &=& \mu \left(a(\tau)\ddot a(\tau) - b(\tau)\ddot b(\tau)\right) \label{g4_t}
\end{eqnarray}
Further note that the sign of $g_3(\tau)$ is inverted in comparison with \cite{Flrespop}.
We can now calculate the transition matrix elements $D_l^{nm}$ (\ref{Eq:FT_diab_coupling}) and $F_l^{nm}$ (\ref{Eq:FT_pert_coupling}):
\begin{eqnarray}
 D_l^{nm} &=& v_{1,l}^{nm} + v_{2,l}^{nm} \\
 F_l^{nm} &=& v_{3,l}^{nm} + v_{4,l}^{nm}
\end{eqnarray}
\begin{eqnarray}
 v_{1,l}^{nm} &=& - \dot \imath \int_0^1 d\tau e^{2 \pi \dot \imath l \tau} e^{-\frac{2 \pi \dot \imath}{\omega} \Delta\nu_{mn}(\tau)}
\left(r - \frac{1}{r}\right) \frac{\dot r}{r} \frac{\Braket{n;r|\hat{f}^1|m;r}}{q_n(r) - q_m(r)} \nonumber\\
\label{pert_zener_1}\\
\nonumber\\
 v_{2,l}^{nm} &=& - \dot \imath \int_0^1 d\tau e^{2 \pi \dot \imath l \tau} e^{-\frac{2 \pi \dot \imath}{\omega} \Delta\nu_{mn}(\tau)}
\left(r + \frac{1}{r}\right) \frac{\dot r}{r} \frac{\Braket{n;r|\hat{f}^3 + \hat{f}^5|m;r}}{q_n(r) - q_m(r)} \nonumber\\
\label{pert_zener_2}\\
\nonumber\\
 v_{3,l}^{nm} &=& \frac{1}{2 \pi \hbar} \int_0^1 d\tau e^{2 \pi \dot \imath l \tau} e^{-\frac{2 \pi \dot \imath}{\omega} \Delta\nu_{mn}(\tau)}
g_2(\tau) \Braket{n;r|\hat{f}^2|m;r} \nonumber\\
\label{pert_acc_1} \\
\nonumber\\
 v_{4,l}^{nm} &=& \frac{1}{2 \pi \hbar} \int_0^1 d\tau e^{2 \pi \dot \imath l \tau} e^{-\frac{2 \pi \dot \imath}{\omega} \Delta\nu_{mn}(\tau)}
g_4(\tau) \Braket{n;r|\hat{f}^4 + \hat{f}^6|m;r}  \ \textrm{.} \nonumber \\ 
\label{pert_acc_2}
\end{eqnarray}
After calculating these quantities and diagonalizing $M(r)$ and $H_M$, it is straightforward to reproduce the theoretical predictions contained in this work. 

\section{Theoretical predictions}
\label{App:theoretical_predictions}

\begin{table}[htp]
\begin{center}
\begin{tabular}{|c | c | c | c | c |}
    \hline
    $\omega_{res}^{n4,l}$ & $\tau_{\textrm{int}}$ & $\tau_{\textrm{low}}$ & state $n$ & order $l$ \\ \hline \hline
    3.966 & 304 & 0.181 & 1 & 3\\ \hline
    5.030 & 328 & 0.482 & 7 & 2\\ \hline
    5.122 & 1014 & 0.493 & 10 & 4\\ \hline
    5.944 & 39.4 & 0.271 & 1 & 2\\ \hline
    6.829 & 133 & 0.657 & 10 & 3\\ \hline
    7.720 & 1098 & 0.743 & 13 & 4\\ \hline
    10.09 & 40.5 & 0.970 & 7 & 1\\ \hline
    10.24 & 17.3 & 0.985 & 10 & 2\\ \hline
    10.29 & 144 & 0.990 & 13 & 3\\ \hline
    11.84 & 4.91 & 0.540 & 1 & 1\\ \hline
    15.11 & 1647 & 1.77 & 22 & 4\\ \hline
    15.44 & 18.7 & 1.48 & 13 & 2\\ \hline
\end{tabular}
\end{center}
\caption{Numerical values for all predicted resonance frequencies between the values 0 and 16 with an interaction time $\tau_{\textrm{int}}$ less than 2000. 
Information is provided on the corresponding lower threshold $\tau_{\textrm{low}}$, the quantum number of the coupling instantaneous eigenstate $n$
and the photon process order $l$ of the resonance.}
\label{Tab:ratio_resonances}
\end{table}

\begin{table}[htp]
\begin{center}
\begin{tabular}{|c | c | c | c | c |}
    \hline
    $\omega_{res}^{n4,l}$ & $\tau_{\textrm{int}}$ & $\tau_{\textrm{low}}$ & state $n$ & order $l$ \\ \hline \hline
    2.584 & 589 & 0.254 & 7 & 4\\ \hline
    3.446 & 150 & 0.339 & 7 & 3\\ \hline
    3.972 & 309 & 0.178 & 1 & 3\\ \hline
    5.128 & 1207 & 0.504 & 10 & 4\\ \hline
    5.170 & 40.0 & 0.508 & 7 & 2\\ \hline
    5.954 & 39.9 & 0.268 & 1 & 2\\ \hline
    6.836 & 151 & 0.672 & 10 & 3\\ \hline
    7.807 & 361 & 0.728 & 13 & 4\\ \hline
    10.25 & 18.8 & 1.01 & 10 & 2\\ \hline
    10.35 & 11.0 & 1.02 & 7 & 1\\ \hline
    10.41 & 63.7 & 0.970 & 13 & 3\\ \hline
    11.86 & 4.90 & 0.534 & 1 & 1\\ \hline
    13.18 & 1284 & 3.29 & 20 & 4\\ \hline
    15.21 & 1019 & 1.87 & 22 & 4\\ \hline
    15.62 & 11.2 & 1.44 & 13 & 2\\ \hline
\end{tabular}
\end{center}
\caption{Analogue to Tab. \ref{Tab:ratio_resonances} but for the breathing driving law.}
\label{Tab:breathing_resonances}
\end{table}

\begin{table}[htp]
\begin{center}
\begin{tabular}{|c | c | c | c | c |}
    \hline
    $\omega_{res}^{n4,l}$ & $\tau_{\textrm{int}}$ & $\tau_{\textrm{low}}$ & state $n$ & order $l$ \\ \hline \hline
    0.3682 & 1985 & 0.0367 & 7 & 3\\ \hline
    0.3823 & 1889 & 0.0381 & 7 & 2\\ \hline
    0.3976 & 1975 & 0.0397 & 7 & 4\\ \hline
    0.4733 & 1176 & 0.0472 & 7 & 2\\ \hline
    0.4970 & 842 & 0.0496 & 7 & 3\\ \hline
    0.5232 & 741 & 0.0522 & 7 & 4\\ \hline
    0.5522 & 712 & 0.0551 & 7 & 1\\ \hline
    0.5847 & 525 & 0.0583 & 7 & 2\\ \hline
    0.6213 & 303 & 0.0620 & 7 & 3\\ \hline
    0.6627 & 185 & 0.0661 & 7 & 1\\ \hline
    0.7100 & 124 & 0.0708 & 7 & 4\\ \hline
    0.7647 & 88.6 & 0.0763 & 7 & 2\\ \hline
    0.8284 & 63.7 & 0.0826 & 7 & 3\\ \hline
    0.9037 & 45.0 & 0.0902 & 7 & 2\\ \hline
    0.9941 & 31.3 & 0.0992 & 7 & 4\\ \hline
    1.105 & 22.0 & 0.110 & 7 & 2\\ \hline
    1.243 & 15.7 & 0.124 & 7 & 3\\ \hline
    1.420 & 11.5 & 0.142 & 7 & 4\\ \hline
    1.657 & 8.51 & 0.165 & 7 & 1\\ \hline
    1.989 & 6.42 & 0.199 & 7 & 2\\ \hline
    2.487 & 4.95 & 0.248 & 7 & 3\\ \hline
    3.317 & 3.94 & 0.332 & 7 & 1\\ \hline
    4.982 & 3.31 & 0.500 & 7 & 4\\ \hline
    5.905 & 411 & 0.271 & 1 & 2\\ \hline
    6.228 & 1134 & 0.555 & 13 & 3\\ \hline
    7.795 & 691 & 0.691 & 13 & 2\\ \hline
    8.559 & 875 & 1.98 & 20 & 4\\ \hline
    9.926 & 28.9 & 1.01 & 10 & 2\\ \hline
    10.04 & 2.97 & 1.02 & 7 & 3\\ \hline
    10.28 & 347 & 2.33 & 20 & 4\\ \hline
    10.42 & 218 & 0.912 & 13 & 1\\ \hline
    11.74 & 5.06 & 0.538 & 1 & 2\\ \hline
    12.86 & 136 & 2.83 & 20 & 3\\ \hline
\end{tabular}
\end{center}
\caption{Analogue to Tab. \ref{Tab:ratio_resonances} but for the volume-preserving driving law.}
\label{Tab:vol_resonances}
\end{table}

\end{document}